\documentstyle[epsfig]{elsart}

\newcommand \be{\begin{equation}}
\newcommand \ba{\begin{eqnarray}}
\newcommand \ee{\end{equation}}
\newcommand \ea{\end{eqnarray}}

\begin{document}
\runauthor{Zhou and Sornette}
\begin{frontmatter}
\title{Renormalization group analysis of the 2000-2002
anti-bubble in the US S\&P 500 index: Explanation of the hierarchy
of 5 crashes and prediction}
\author[iggp]{\small{Wei-Xing Zhou}},
\author[iggp,ess,nice]{\small{Didier Sornette}\thanksref{EM}}
\address[iggp]{Institute of Geophysics and Planetary Physics, University of California
Los Angeles, CA 90095-1567}
\address[ess]{Department of Earth and Space Sciences, University of California
Los Angeles, CA 90095-1567}
\address[nice]{Laboratoire de Physique de la Mati\`ere Condens\'ee,
CNRS UMR 6622 and Universit\'e de Nice-Sophia Antipolis, 06108 Nice Cedex 2, France}
\thanks[EM]{Corresponding author. Department of Earth and Space
Sciences and Institute of Geophysics and Planetary Physics,
University of California, Los Angeles, CA 90095-1567, USA. Tel:
+1-310-825-2863; fax: +1-310-206-3051. {\it E-mail address:}\/
sornette@moho.ess.ucla.edu (D. Sornette)\\
http://www.ess.ucla.edu/faculty/sornette/}
\begin{abstract}
We propose a straightforward extension of our previously proposed
log-periodic power law model of the ``anti-bubble'' regime of the
USA market since the summer of 2000, in terms of the
renormalization group framework to model critical points. Using a
previous work by Gluzman and Sornette (2002) on the classification
of the class of Weierstrass-like functions, we show that the five
crashes that occurred since August 2000 can be accurately modelled
by this approach, in a fully consistent way with no additional
parameters. Our theory suggests an overall consistent organization
of the investors forming a collective network which interact to
form the pessimistic bearish ``anti-bubble'' regime with
intermittent acceleration of the positive feedbacks of pessimistic
sentiment leading to these crashes. We develop retrospective
predictions, that confirm the existence of significant arbitrage
opportunities for a trader using our model. Finally, we offer a
prediction for the unknown future of the US S\&P500 index
extending over 2003 and 2004, that refines the previous prediction
of Sornette and Zhou (2002).
\end{abstract}
\end{frontmatter}

\section{Introduction}
\label{sec:intro}

In two recent papers \cite{SZ02,AntiBubb}, we have analyzed 39
world stock market indices from 2000 to the end of 2002 and have
found that 22 of them are in an ``anti-bubble'' regime, defined as
a self-fulfilling decreasing price created by positive
price-to-price feedbacks feeding an overall pessimism and negative
market sentiments further strengthened by interpersonal
interactions. Mathematically, we characterize anti-bubbles by a
power law decrease of the price (or of the logarithm of the price)
as a function of time decorated by decelerating/expanding
log-periodic oscillations. We have used several parametric and
non-parametric statistical tools which all confirm the very high
significance of the reported mathematical signature of herding
characterizing these bearish anti-bubble phases. In addition, the
majority of European and Western stock market indices as well as
other stock indices have been found to exhibit practically the
same log-periodic power law anti-bubble structure as found for the
USA S\&P500 index. These anti-bubbles are found to start
approximately at the same time, August 2000, in all these markets
and show a remarkable degree of synchronization worldwide,
suggesting that the descent of the worldwide stock markets since
2000 is an international event. These results augment the number
of anti-bubble cases in addition to the previous examples
documented on the Japanese, Gold and Russian markets
\cite{Nikkei99,JSanticonfirm}.

These analyses \cite{SZ02,AntiBubb} are based on a general theory
of financial crashes and of stock market instabilities developed
in a series of works by D. Sornette and his co-workers over the
last decade
\cite{SJB96,CriCrash99,JohSorLed99,CriCrash00,SorJoh01,bookcrash}.
The main ingredient of the theory is the existence of positive
feedbacks, that is, self-reinforcements, which leads to
cooperative herding and imitation between investors. As a
consequence, the actions of investors tend to produce waves of
contrariant and imitative behaviors leading to self-reinforcing
bullish or bearish phases, themselves decorated by the opposite
regimes in a hierarchical fashion. Positive feedbacks, when
unchecked, can produce runaways until the deviation from the
fundamental price is so large that other effects can be abruptly
triggered and lead to rupture or crashes. Alternatively, it can
lead to prolonged depressive bearish markets for most of the
developed stock markets in the world \cite{AntiBubb}.

Figure \ref{FigWeiPrice} shows the USA S\&P 500 index from August
21, 2000 to August 24, 2002 and the fits by the log-periodic power
law formula
\begin{equation}
\ln [p\left(t\right)] \approx A + B (t-t_c)^{\alpha} +C
(t-t_c)^\alpha \cos\left[ \omega \ln \left( t-t_c \right) -\phi
\right]~, \label{Eq:fit1}
\end{equation}
(where $t_c$ is the starting time of the anti-bubble) including a
single angular log-frequency $\omega$ (continuous line) as well as
its harmonics $2 \omega$ (dashed line). These two theoretical
curves constituted the main results of our previous work
\cite{SZ02} which led us to issue a prediction for the future
evolution of the US market (see also our web site for monthly
updates of our predictions). Notwithstanding the remarkable
capture of the overall decay and of the expanding alternation of
ups and downs, the five crashes outlines by the arrows are not
accounted by this theory. These five crashes constitute an
important characteristic of this time series which is much rougher
and singular than predicted by the gentle cosine function used in
expression (\ref{Eq:fit1}). The importance of these five crashes
can be seen both at the level of the geometrical description of
the price time series and with respect to their meaning and
implications in the overall organization of the social network of
investors.

Our purpose is to propose a simple extension of expression
(\ref{Eq:fit1}) based on a renormalization group analysis of the
critical behavior embodied by (\ref{Eq:fit1}) that accounts very
accurately for the presence of these five crashes. The theory
developed here shows that expression (\ref{Eq:fit1}) is nothing
but the first Fourier component of a more complete infinite
Fourier series expansion of the general critical log-periodic
power law behavior. This extension rationalizes the hierarchical
geometric series of the time occurrences of these five crashes.
Within our new theory, the qualitative explanation for the
occurrence of these five crashes is similar to that underlying
crashes following speculative bubbles. The main qualitative
ingredient of the theory is the existence of positive feedbacks,
of the kind which has been documented to occur in stock markets as
well as in the economy. Positive feedbacks, i.e.,
self-reinforcement, refer to the fact that, conditioned on the
observation that the market has recently moved up (respectively
down), this makes it more probable to keep it moving up
(respectively down), so that a large cumulative move may ensue. In
a dynamical model, positive feedbacks with nonlinear
amplifications have been shown to give rise to finite-time
singularities which are good mathematical representations of the
market price trajectory on the approach to crashes
\cite{Ide1,Ide2}. However, such dynamical formulation describes
only the price trajectory up to its singularity. It does not
describe the rebound and ensuing evolution after the singularity.
In this sense, our present approach is more powerful because it
shows a deep connection between the existence and organization of
the five crashes on the one hand and the overall log-periodic
power law decay of the market crash on the other hand.

In an anti-bubble such as the regime of the S\&P500 since August
2000, after a bottom, the positive feedbacks described above lead
to the following qualitative scenario. First, the price rebounds
as investors expect that the bad times may be behind them.
However, the euphoria is short-lived as the fundamentals and the
overall market sentiment are negative, leading to a slowdown and
even reversal of the positive into a negative trend. The negative
trend is then amplified by the collective herding behavior of
investors, who, similarly to lemmings rushing over a cliff, run to
sell, leading to each of these five observed crashes. Within our
framework, these five crashes pertaining to the S\&P500
anti-bubble phase are thus intrinsically endogenous to the market
dynamics, similarly to most of the crashes found in many markets
worldwide \cite{EndoExo}.

This paper is organized as follows. In Sec.~\ref{S:derive}, we
summarize the main concepts and results of a renormalization group
theory of the anti-bubble regime, leading to solutions taken the
form of Weierstrass-type functions \cite{GS02}. We present our
fitting procedure in Sec.~\ref{S2:fitting} and the determination
of the relevant number of terms in our model in Sec.~\ref{S2:N}.
The class of models describing the discretely scale invariant
hierarchy of crashes is presented in Sec.~\ref{S2:Mimick}. Its
variants leading to non-singular price trajectories are
investigated in Sec.~\ref{S2:phase}. In Sec.~\ref{S:predict}, we
then use three variants with different choices of the phases of
the log-periodic components to offer a new analysis of the
prediction of the S\&P 500 index in the coming year. This set of
predictions extend and refine our previous prediction \cite{SZ02}.
Sec.~\ref{S:concl} concludes.

\section{Construction of Weierstrass-type functions}
\label{S:derive}

$\ln[p\left(t\right)]$ given by expression (\ref{Eq:fit1}) is
endowed with the symmetry of discrete scale invariance
\cite{DSI98}. Indeed, multiplying $t-t_c$ by the factor $1/\gamma
= e^{2\pi/\omega}$, we have \be \ln [p\left((t-t_c)/\gamma
\right)] =  A {\mu-1 \over \mu} ~+~{1 \over \mu} \ln [p\left(t-t_c
\right)]~, ~~{\rm with}~~\mu = \gamma^{\alpha}~. \label{gmjkerkl}
\ee Equation (\ref{gmjkerkl}) expresses the fact that the prices,
at two times related by a simple contraction with factor $\gamma$
of the interval since the beginning of the anti-bubble, are
related themselves by a simple contraction with factor $\mu$ (up
to the additive term $A (\mu-1 )/\mu$). The scale invariance
equation (\ref{gmjkerkl}) is a special case of a general
one-parameter renormalization group equation, which captures the
collective behavior of investors in speculative regimes (see
\cite{bookcrash} for a pedagogical introduction to the ideas of
the renormalization group as applied to finance). Using the
renormalization group (RG) formalism on the stock market index $p$
amounts to assuming that the index at a given time $t$ is related
to that at another time $t'$ by the general transformations
(\ref{firstt}) and (\ref{eq:fK}) below, first proposed for the
stock market in Ref.~\cite{SJB96} on the variable \be
{\mathcal{F}}(x)=\ln[p(t)]- \ln[p(t_c)]~, \ee such that $F=0$ at
the critical point at $t_c$. The first expression on the distance
$x=|t-t_c|$ to the critical point occurring at $t_c$ reads \be x'
= R(x) ~, \label{firstt} \ee where $R$ is called the RG flow map.
The second expression describes the corresponding transformation
of the price and reads
\begin{equation}
{\mathcal{F}}(x) = {\mathcal{G}}(x) +
\frac{1}{\mu}{\mathcal{F}}[R(x)]~. \label{eq:fK}
\end{equation}
The constant $\mu>1$ describes the scaling of the index evolution
upon a rescaling of time (\ref{firstt}). The function
${\mathcal{G}}(x)$ represents the non-singular part of the
function ${\mathcal{F}}(x)$ which describes the degrees of freedom
left-over by the decimation procedure involved in the change of
scale of the description. We assume as usual \cite{Goldenfeld}
that the function ${\mathcal{F}}(x)$ is continuous and that $R(x)$
is differentiable.

These equations (\ref{eq:fK}) with (\ref{firstt}) can be solved
recursively to give
\begin{equation}
{\mathcal{F}} = \sum_{n=0}^\infty \frac{1}{\mu^n}
{\mathcal{G}}\left[R^{(n)}(x)\right]~, \label{eq:fKrecur}
\end{equation}
where $R^{(n)}$ is the $n$th iterate of the map $R$. To make further
progress, one expands $R(x)$ as a Taylor series in powers of $x$. For our
purpose, it is sufficient to keep only the first-order
linear approximation
\be
R(x) = \gamma x
\ee
of the map $R$ around the fixed point $R(x_c)=x_c=0$. For the effect
of the next order term proportional to $x^2$, see \cite{Derrida}.
Equation (\ref{eq:fKrecur}) then becomes
\begin{equation}
{\mathcal{F}} = \sum_{n=0}^\infty \frac{1}{\mu^n}
{\mathcal{G}}\left[\gamma^n(x)\right]~. \label{eq:fKlin}
\end{equation}
For the special choice ${\mathcal{G}}(x) = \cos(x)$, expression
(\ref{eq:fKlin}) recovers the
Weierstrass function, famous for the fact that it was proposed by Weierstrass
in 1876 as the first example of a continuous function which is no-where
differentiable. This opened up the field of ``fractals'' \cite{Edgar}.

Following Ref.~\cite{GS02}, the infinite series in (\ref{eq:fKlin}) can be expressed
in a more convenient way
by (1) applying the Mellin transform to (\ref{eq:fKlin}),
(2) re-expanding by ordering with respect to the poles
of the Mellin transform of ${\mathcal{F}}$ in the complex plane, and (3)
perform the inverse Mellin transform to obtain an expansion for
${\mathcal{F}}$ in terms of power laws expressing the critical behavior of the
solution of the RG equation (\ref{eq:fK}). This set of operation amounts
to a resummuation of the infinite series in (\ref{eq:fKlin}) which is better
suited to describe the critical behavior and its leading corrections to scaling.
The result is that ${\mathcal{F}}(x)$ can be expressed as the sum
of a singular part ${\mathcal{F}}_s(x)$ and of a regular part
${\mathcal{F}}_r(x)$ \cite{GS02}:
\begin{equation}
{\mathcal{F}}(x) = {\mathcal{F}}_s(x) + {\mathcal{F}}_r(x)~,
\label{eq:Fsr}
\end{equation}
where
\begin{equation}
{{\mathcal{F}}}_s(x) = \sum_{n=0}^\infty A_n x^{-s_n}~,
\label{eq:Fs}
\end{equation}
with
\begin{equation}
A_n = \frac{\hat{{\mathcal{G}}}(s_n)}{\ln\gamma}~, \label{eq:An}
\end{equation}
and
\begin{equation}
s_n = -m+i\frac{2\pi}{\ln\gamma}n~,~~{\rm with}~~m={\ln \mu \over \ln \gamma}~.
\label{eq:sn}
\end{equation}
The coefficients $A_n$ control the relative weights of
the log-periodic corrections (resulting from the
imaginary parts of the complex exponents $s_n$) to the leading power law behavior
described by the first term $n=0$.
The regular part ${\mathcal{F}}_r(x)$ does not exhibit any singularity
at $x=0$ or anywhere else and will not be considered further.

As an illustration, for the Weierstrass case with
${\mathcal{G}}(x) = \cos(x)$, for large $n$, the coefficients $A_n$ are given by
\begin{equation}
 A_n = \frac{1}{\ln\gamma}\frac{1}{n^{m+0.5}}{\rm{e}}^{i\psi_n}~,
{\rm with}~~\psi_n =\omega n\ln(\omega n)~~{\rm and}~~ \omega = {2\pi \over \ln \gamma}~.
 \label{eq:Ancos}
\end{equation}
More generally, Ref.~\cite{GS02} has shown that, for a large class
of systems, the coefficients $A_n$ of the power expansion of the
singular part ${\mathcal{F}}_s(x)$ can be expressed as the product
of an exponential decay by a power prefactor and a phase \be A_n =
{1 \over \ln \gamma}~~ {1 \over n^p}~e^{- \kappa n}~e^{i
\psi_n},~~~~{\rm for~large}~~n~,  \label{classangener} \ee where
$p$, $\kappa  \geq 0$ and $\psi_n$ are determined by the form of
${\mathcal{G}}(x)$ and the values of $\mu$ and $\gamma$. Systems,
with quasi-periodic ${\mathcal{G}}(x)$ and/or such that
${\mathcal{G}}(x)$ has compact support, have coefficients $A_n$
decaying as a power law $A_n \sim n^{-p}$ ($\kappa = 0$) leading
to strong log-periodic oscillatory amplitudes. If in addition, the
phases of $A_n$ are ergodic and mixing, the observable presents
singular properties everywhere, similar to those of
``Weierstrass-type'' functions (\ref{eq:Ancos}).

In the following, we restrict our attention to the
``Weierstrass-type'' class of models with $\kappa=0$ and extend
the log-periodic power law formula (\ref{gmjkerkl}) for the fit of
empirical price time series by using the class of models
${{\mathcal{F}}}_s(x)$ given by (\ref{eq:Fs}) with the
parametrization (\ref{classangener}) with $\kappa=0$.
Specifically, we shall use the following function for the fit of
price time series:
\begin{equation}
{{\mathcal{F}}}(x) = A + B x^{m} + C \sum_{n=1}^N
n^{-m-0.5}{\rm{e}}^{i\psi_n}x^{-s_n}~, \label{eq:fx}
\end{equation}
where $A, B$ and $C$ are three parameters and
$N$ is the number of log-periodic harmonics kept
in the description. In the second term of the right-hand-side of
(\ref{eq:fx}), we have used $s_{n=0}=-m$ as seen from (\ref{eq:sn}).
We shall play with different
parameterizations for the phases $\psi_n$, which can
have forms different from
$\psi_n =\omega n\ln(\omega n)$ given in (\ref{eq:Ancos})
for the Weierstrass function.
We shall also test the robustness of the fits with (\ref{eq:fx})
by using the more general function
\begin{equation}
\ln [p(t)] = A + B x^{m} + {\Re{\left(\sum_{n=1}^N C_n
{\rm{e}}^{i\psi_n}x^{-s_n}\right)}}~, \label{Eq:lnpt}
\end{equation}
where ${\Re{()}}$ stands for the real part. Eq.~(\ref{eq:fx}) is
recovered for $C_n = C/n^{m+0.5}$.

\section{Description of the S\&P500 anti-bubble by multiple singularities}
\label{S:sing}

\subsection{Fitting procedure} \label{S2:fitting}

The USA S\&P 500 daily price series of the 2000 anti-bubble from
August 21, 2000 to August 24, 2002, as shown in
Fig.~\ref{FigWeiPrice}, is fitted with expression (\ref{Eq:lnpt})
by using a standard mean-square minimization procedure. We exploit
the linear dependence of $\ln [p(t)]$ on the parameters $A, B$ and
$C_n$ as follows. Rewriting formula (\ref{Eq:lnpt}) as
\begin{equation}
y(x) = A + B f(x;\Phi) + \sum_{n=1}^N C_n g_n(x;\Phi)~,
\label{eq:yt}
\end{equation}
where $\Phi$ encapsulates all the non-linear parameters,
given the data set $\{(x_j, y_j):j=1,...,J\}$, the function to
minimize is
\begin{equation}
Q(A,B,C_n; \Phi) = \sum_{j=1}^J [y_j-y(x_j)]^2~. \label{eq:obj}
\end{equation}
Following \cite{JohSorLed99,CriCrash00}, the $N+2$ linear
parameters $A$, $B$ and $C_n$ are slaved to the nonlinear parameters
$\Phi$ by solving the following linear equations:
\begin{equation}
\partial Q(A,B,C_n; \Phi) / \partial (A,B,C_n) = 0~.
\label{eq:dQ}
\end{equation}
Let $\vec{A} = [\vec{I}, \vec{f}, \vec{g}_1, \cdots, \vec{g}_N]$,
where $\vec{I}$ is the identity column vector of $J$ elements,
$\vec{f} = [f(x_1), f(x_2), \cdots, f(x_J)]^T$ and $\vec{g}_n =
[g_n(x_1), g_n(x_2), \cdots, g_n(x_J)]^T$. Then the solution of
Eq.~(\ref{eq:dQ}) is
\begin{equation}
[A,B,C_1,\cdots,C_N]^T = (\vec{A}^T \vec{A})^{-1}
(\vec{A}\vec{y})^T~, \label{eq:solution}
\end{equation}
where $\vec{y} = [y_1, y_2, \cdots, y_J]^T$.
In the minimization of
$Q(A,B,C_n; \Phi)$ given by (\ref{eq:obj}), the linear parameters
$A$, $B$ and $C_n$ are determined accordingly to
Eq.~(\ref{eq:solution}) as a function of the nonlinear parameters
$\Phi$. The minimization process then reduces to minimizing $Q(A,B,C_n; \Phi)$
with respect only to the nonlinear parameters $\Phi$.
In order to obtain the global solution,
we employ the taboo search \cite{Taboo} to determine an ``elite
list'' of solutions as the initial solutions of the ensuing line
search procedure in conjunction with a quasi-Newton method. The
best fit is regarded to be globally optimized.

\subsection{Jack-knife method to truncate the infinite series}
\label{S2:N}

For a practical implementation, it is necessary to truncate the
infinite series in (\ref{Eq:lnpt}) and (\ref{eq:yt}) at a finite order $N$.
The larger $N$ is, the finer are the structures, as seen from the fact that,
using (\ref{eq:sn}),
the real part of $x^{-s_n}$ equal to $x^m \cos\left[
\omega n \ln (t-t_c)\right]$  involves faster and faster oscillations as $n$ increases.
Using a larger number $N$ of terms in (\ref{Eq:lnpt}) and (\ref{eq:yt})
is bound to improve the quality of the fit by construction since
the mean-square minimization with $N+1$ terms is obviously encompassing
the mean-square minimization with $N$ terms, this later case being
retrieved by imposing $C_{N+1}=0$. We thus need a criterion for characterizing
the trade-off between the improvement of the fits on one hand when
increasing $N$ and
the associated increase in complexity (or lost of parsimony).
For this, we adapt the Jack-knife method \cite{knife74} in the following way.
For a fixed $N$, we take away one observation
$(x_k, y_k)$ from the time series $\{(x_j,y_j): j=1,\cdots,J\}$, estimate
the best fit characterized by its nonlinear parameters
$\Phi_{k}$ using all observations but $(x_k, y_k)$, and
try to predict $\hat{y}_k$, using the estimated function. We then
repeat this procedure for every observation
and get the average ``prediction error''
\begin{equation}
\sigma_N = \left[\frac{1}{J} \sum_{j=1}^J (y_j - \hat{y}_j)^2\right]^{1/2}~,
\end{equation}
which is a function of $N$. The optimal $N$, noted $N^*$, is such that by
\begin{equation}
\sigma_{N^*} = \min_N \{\sigma_N: N\in{\mathcal{N}}\}~. \label{eq:N}
\end{equation}
This Jack-knife method is used usually for a sample of large size,
generated by an unknown but fixed regression curve (e.g. a
polynomial of an unknown degree). By the Jack-knife method, one
determines once and for all the preferable degree of the
approximating polynomial for the whole sample, i.e., the
regression curves do not change from one sample point to another
sample point. If the size of the time series $J$ is too large,
$\sigma_N$ tends asymptotically to the the r.m.s. $\chi_N$ of the
fit errors in the jack-knife test which tends to the r.m.s. $\chi$
of the fit errors with no points deleted. Hence, $\sigma_N$ is a
monotonously decreasing function of $N$. Since we are interested
in predictions over more than a one day horizon to minimize the
impact of market noise, we extend the Jack-knife by partitioning
the time series into groups of $u$ successive points. In each fit,
a group of points are deleted and the previous Jack-knife method
is applied simultaneously to the $u$ consecutive points
to determine which $N$ leads to their best and most robust
prediction.

We fit the S\&P500 time series by Eq.~(\ref{Eq:lnpt}) for $N = 1,
\cdots, 10$. This process is repeated for $u =1, \cdots, 5$. The
set of free nonlinear parameters is $\Phi_N = (t_c, m, \omega,
\psi_1, \cdots,\psi_N)$. In all these fits, we find a very robust
value for the fitted critical time (which is the theoretical
starting time of the 2000 anti-bubble) given by
$t_c={\rm{August~8,~ 2000}} \pm 10$ days. This result is in good
agreement our previous result   $t_c={\rm{August~9\pm 5,~2000}}$
obtained by varying the beginning of the fitted time window
\cite{SZ02}.

\begin{figure}
\begin{center}
\epsfig{file=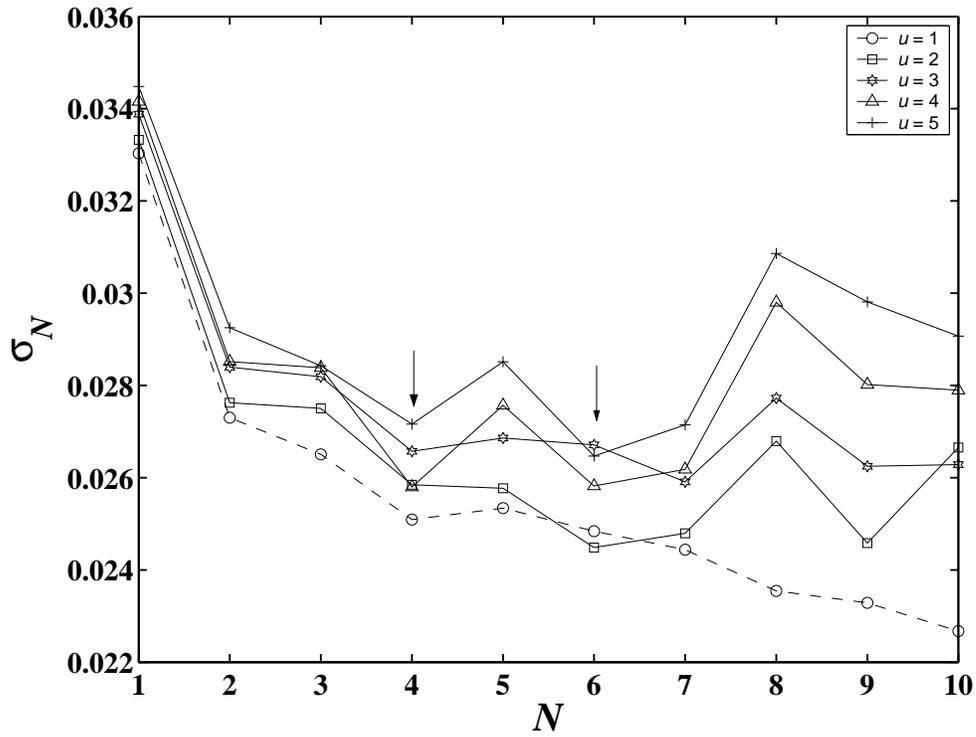,width=13cm}
\end{center}
\caption{Standard deviation $\sigma_N$ of the errors in predicting
the values of $u$ successive data points as a function of the
number $N$ of terms in the series (\ref{eq:fx}) defining the model
used to fit the data. This constitutes a slight generalization of
the Jack-knife method. The downward pointing arrows indicate the
first two minima of $\sigma_N$ at $N=4$ and $N=6$ for $u>1$. The
``global'' minimum lies at $N=6$.} \label{FigWeiJack}
\end{figure}

Figure \ref{FigWeiJack} illustrates how we determine the optimal
value for the number $N$ of terms in (\ref{Eq:lnpt}) using the
extended Jack-knife method describe above. The dashed line with
open circles corresponds to deleting only one point ($u=1$). In
contrast with the monotonous decrease of $\sigma_N$ obtained for
$u=1$, two local minima at $N=4$ and $N=6$ occur for all $u>1$,
while the best minimum of $\sigma_{N^*}$ defined by (\ref{eq:N})
is obtained for $N=6$. We shall thus perform all our fits with six
terms in the series (\ref{Eq:lnpt}) and (\ref{eq:yt}).

\subsection{A parsimonious model of the discrete hierarchy of five crashes} \label{S2:Mimick}

Within the present RG theory whose solution is expressed in the
form of the ``Weierstrass-type functions'' (\ref{eq:fKlin}),
additional singularities other than that occurring by definition
at $x_c=0$ appear when the phases $\psi_n$ in (\ref{classangener})
all vanish so that all log-periodic power laws $x^{-s_n}$ in the
expansion (\ref{eq:Fs}) are perfectly in phase for $x$ coincident
with the infinitely discrete sets $x_q=1/\gamma^q$ with $q$
integer. Appendix B of Ref.~\cite{GS02} demonstrates using a
renormalization group method applied to the series (\ref{eq:Fs})
that ${\mathcal{F}}_S(x)$ behaves, in the vicinity of these
singular points $x_q$, as
\be
{\mathcal{F}}_S(x) \sim {\rm
constant} + c |x-x_q|^{m- {1 \over 2}}~,
\label{mgkjlor}
\ee
where
$c$ is a constant. Thus, for $0 < m < 1/2$, spikes develop that
correspond to a divergence of ${\mathcal{F}}_S(x)$ as $x \to x_q =
1/\gamma^q$. For $1/2 < m < 3/2$, ${\mathcal{F}}_S(x)$ goes to a
finite value as $x \to x_q$ but with an infinite slope (since $0 <
m-{1 \over 2} < 1$). The appearance of this set of singularities
described by (\ref{mgkjlor}) is fundamentally based on the
coherent ``interferences'' between all log-periodic oscillations
in the infinite series expansion (\ref{eq:Fs}). These
singularities disappear as soon as the phases are a function of
$n$ due to the ensuing dephasing. In particular, for generic phase
functions, the phases are ergodic as a function of $n$ and the
singularities give place to no-where differentiable function (for
$m<1$) \cite{GS02}.

\begin{figure}[t]
\begin{center}
\epsfig{file=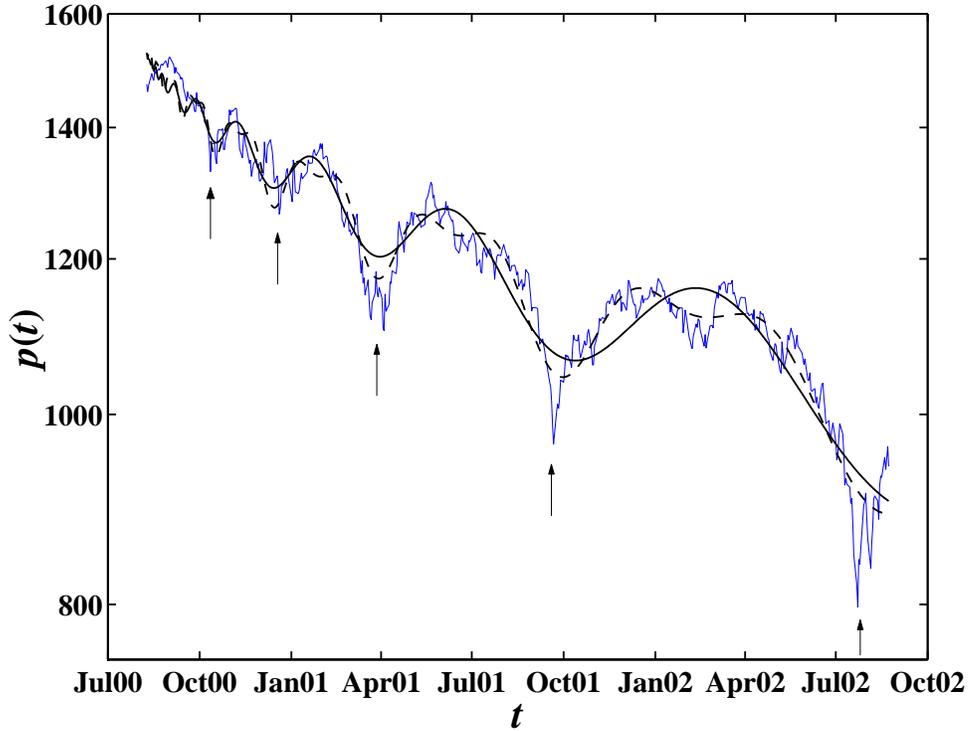,width=13cm}
\end{center}
\caption{The price trajectory of the US S\&P 500 index from August
21, 2000 to August 24, 2002. A clear decaying log-periodic
structure is visible, implying a well developed anti-bubble. The
upward arrows indicate the very large and sharp drops or crashes
in the price evolution, which will be modelled in this paper as
approximate local singularities. The lines are the fits by the
log-periodic power law formula (\ref{Eq:fit1}) including a single
angular log-frequency $\omega$ (continuous line) as well as its
harmonics $2 \omega$ (dashed line).} \label{FigWeiPrice}
\end{figure}

Therefore, in order to reconstruct the five crashes shown
in Figure \ref{FigWeiPrice}, we first impose
$\psi_n =0$ and search for the best fit of the S\&P500 time series by
expression (\ref{eq:fx}) with
$C_n=C/n^{m+0.5}$, $n=1, ...,N$, truncated at $N=6$. Compared with
the more general expression (\ref{Eq:lnpt}), equation
(\ref{eq:fx}) has only three linear
parameters $A$, $B$ and $C$ instead of eight ($A$,
$B$, $C_1$, $\cdots$, and $C_6$). We fit the S\&P500 data from
August 21, 2000 to August 24, 2002 and exclude the
later data (the last four months at the time of writing)
because August 24, 2002 is close to the end of the last
complete log-periodic oscillations and the later data contains
only a small part of the next extrapoled cycle. Our tests show
that taking into account such a fraction of a log-periodic
cycle in such an expanding anti-bubble may
have a significant impact in the quality of the fits. Similarly,
when we fit the time series ending
in March 2002 at which the preceding cycle has completed less than
its period, the retrieved log-periodic structure is quite distorted. We
shall come back to this issue in details in Sec.~\ref{S:predict}
regarding its impact on the prediction of the future.

Figure \ref{FigWeiphase0} shows the best fit (in thick wiggly
line) of the S\&P500 index with formula (\ref{eq:fx}) with
$\psi_n=0$, $C_n=C/n^{m+0.5}$, $n=1, ...,N$, truncated at $N=6$.
The values of the parameters are $t_c= {\rm{July~26, 2000}}$,
$m=0.53$, $\omega=11.43$, $A=7.39$, $B=-0.0141$, and $C=-0.0022$.
The r.m.s. of the fit residuals is $0.0252$ (with only three free
nonlinear parameters). In comparison, the r.m.s. of the fit
residuals with expression (\ref{Eq:fit1}) is $0.0325$ (four free
nonlinear parameters) and with the extension including the second
harmonic is $0.0265$ (five free nonlinear parameters). Since $C$
is determined from the fit with $N=6$, it can be used to
extrapolate to the large $N$ limit and we thus also show the
formula for $N=1000$ (thin smooth line with sharp troughs), which
approximates very well the singularities at the five times of the
crashes. Since $m=0.53$ is larger than $1/2$, the fit selects the
class of singularities with finite end-log(price) and infinite
slope. However, since $m$ is so close to the borderline value
$0.5$, the price trajectory predicted by (\ref{mgkjlor}) is
essentially undistinguishable numerically from a logarithmic
singularity of the form ${\mathcal{F}}_S(x) \sim \ln|x-x_q|$. Such
logarithmic singularities have been derived from a simple theory
of positive feedback on risk aversion \cite{Boucont} and has been
used as a trick to remove the exponent parameter from the fitting
procedure \cite{van1,van2} (see however \cite{JS99manifesto}).

This model reproduces remarkably well the structure of the S\&P500
index, including both the overall log-periodic power law decay of
the anti-bubble and the discrete hierarchy of the five crashes
shown with arrows in Figure \ref{FigWeiPrice}.

\subsection{The role of the phase} \label{S2:phase}

\begin{figure}
\begin{center}
\epsfig{file=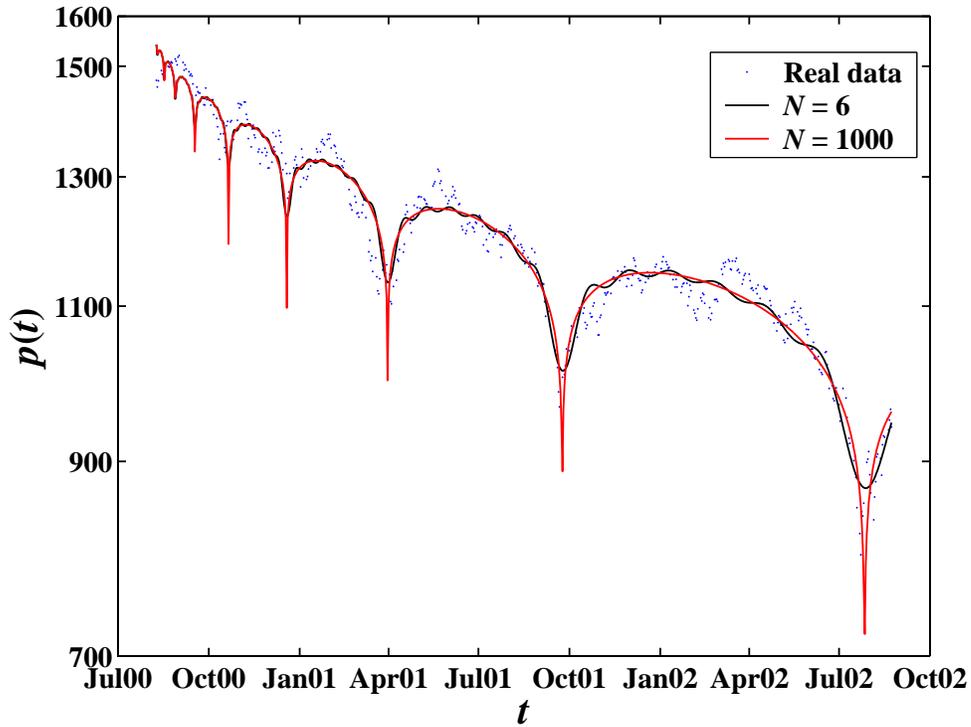,width=13cm}
\end{center}
\caption{Fit of the price trajectory of the US S\&P 500 (dots) by
the model defined by Eq.~(\ref{eq:fx}) with $N=6$ and $\psi_n=0$.
The fitted function is shown as the continuous wiggly line. The
values of the fitted parameters are: $t_c= {\rm{July~26, 2000}}$,
$m=0.53$, $\omega=11.43$, $A=7.39$, $B=-0.0141$, and $C=-0.0022$.
The r.m.s. of the fit residuals is $0.0256$. The smooth continuous
curve exhibiting local singularities (sharp drops) is constructed
by inserting these fitted parameters into Eq.~(\ref{eq:fx}) with
$N=1000$.} \label{FigWeiphase0}
\end{figure}

Ref.~\cite{GS02} has shown that the dependence of the phases
$\psi_n$ in Eq.~(\ref{eq:fx}) and Eq.~(\ref{Eq:lnpt}) as a
function of $n$ plays a central role in determining the structure
of the solution. Expanding $\psi_n$ as $\psi_n = a_p n^p + ... +
a_1 n + a_0$, where $p$ is the degree of the polynomial expansion
of $\psi_n$ in powers of $n$, we find by fitting it to the price
series for $p = 0,1,2,3,4,5$ that only for $p=0$ and $p=1$ can one
obtain the singular behavior obtained for $\psi_n=0$ (recovered
for $p=0$) shown in figure \ref{FigWeiphase0}. Intuitively, the reason is that
only for $p=0$ and $p=1$ are the phases $\psi_n$ sufficiently
regular to allow for constructive interference of the successive
harmonics of Eq.~(\ref{eq:fx}) and Eq.~(\ref{Eq:lnpt}).

Let us now sketch the proof of the following assertion:
the necessary and sufficient
condition for the existence of localized singularities is
\be
\psi_n = a_1n+a_0~. \label{Eq:SufNes}
\ee
The crucial remark is that the cosine part $\cos[n\omega\ln(t-t_c)-\psi_n]$ of the
$n$th harmonic term in Eq.~(\ref{eq:fx}) can always be rewritten as
$\cos[n\omega\ln(\tau)-a_0]$ with $\tau=(t-t_c)/T$ where
$T=\e^{a_1/\omega}$. The coefficient $a_1$ can thus be gauged
away in a change of time units.
Now, it is clear that for $\omega\ln(\tau) =
2 \pi k $ where $k$ is an integer, the cosines
$\cos[n\omega\ln(\tau)-a_0]$ are equal to $\cos(a_0)$ for all $n$.
Thus, the infinite sum of the Weierstrass series is equal to
$\cos(a_0)$ times the sum obtained by putting all phases exactly
equal to zero.
This constitutes the sufficient part of the proof that phases
of the form (\ref{Eq:SufNes}) are equivalent to phases identically
equal to zero with respect to the occurrence of localized singularities
(\ref{mgkjlor}). For
the necessary part of the proof, it is easy to check by
contradiction that a power $n^p$ with $p>1$ will give a moving
amplitude for each harmonics thus breaking down the synchronization
of the phases for the harmonics. One sees that the zero phase
condition $\psi_n=0$ of the existence of localized singularities
\cite{GS02} is a special case of the linear phase condition
(\ref{Eq:SufNes}).

For the fit with $p=1$, that is with expression (\ref{Eq:SufNes}),
we find the following parameter values:
$t_c = {\rm{July~31, 2000}}$, $m=0.54$, $\omega=11.35$,
$a_1=49.56$, $a_0= 3.35$, $A=7.37$, $B=-0.0125$, $C= 0.0020$, with
a r.m.s. of the fit residuals equal to $\chi=0.0249$. We find that
the parameters $t_c$, $m$, $\omega$, $A$, $B$ and $C$ are very close to
those obtained with the fit shown
Fig.~\ref{FigWeiphase0}, indicating a good robustness with
respect to the addition of the two parameters $a_0$ and $a_1$.
Note also the very small improvement of the
r.m.s. obtained by using (\ref{Eq:SufNes}) compared with the
r.m.s. of $0.0252$ obtained by imposing $a_0=a_1=0$ and shown in
Figure \ref{FigWeiphase0}. As the addition of these two parameters
$a_0$ and $a_1$ provides only a very minor improvement, it would
not be qualified by using an Aikaike or information entropic
criterion. In our investigation below on the potential of this
formulation for prediction, we shall thus focus on the more
parsimonious specification with zero phases $\psi_n=0$. Other
tests not presented here confirm the over-determination of the
Weierstrass-like formula with the two parameters $a_0$ and $a_1$
because it gives very poor prediction skills in comparison with
the simpler model $\psi_n=0$.

\begin{figure}
\begin{center}
\epsfig{file=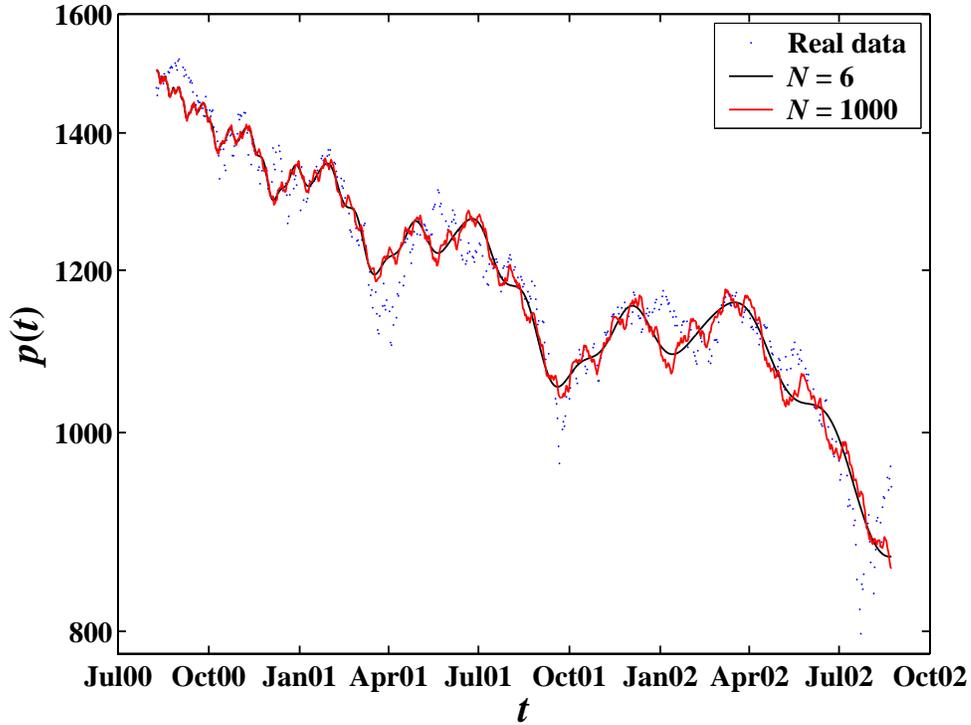,width=13cm}
\end{center}
\caption{Fit of the price trajectory of the US S\&P 500 (dots)
using Eq.~(\ref{eq:fx}) with $N=6$ and $\psi_n=\omega n^2$. The
values of the parameters are: $t_c= {\rm{August~7, 2000}}$,
$m=0.69$, $\omega=10.61$, $A=7.32$, $B=-0.0043$, and $C= 0.0008$.
The r.m.s. of the fit residuals is $0.0312$. The thin wiggly curve
is constructed by inserting the fitted parameters into
Eq.~(\ref{eq:fx}) with $N=1000$. Freeing the parameter $\omega$ in
the phase $\psi_n=\omega n^2$ by adopting $\psi_n = a n^2$
improves the goodness-of-fit to the r.m.s. of fit residuals equal
to $\chi=0.274$.}\label{FigWeiWn2}
\end{figure}

To demonstrate the phenomenon of destructive interferences between
the log-periodic oscillations of successive harmonics, let us
consider the simple model $\psi_n = \omega n^2$. Following the
same fitting procedure as in Sec.~\ref{S2:Mimick} with $N=6$,
Fig.~\ref{FigWeiWn2} shows the best fit (smooth continuous line)
compared with the data (dots), corresponding to the parameters
$t_c= {\rm{August~7, 2000}}$, $m=0.69$, $\omega=10.61$, $A=7.32$,
$B=-0.0043$, and $C= 0.0008$. The r.m.s. of the fit residuals is
$0.0312$ (three free nonlinear parameters).
The wiggly continuous curve is constructed by inserting
these fitted parameters into Eq.~(\ref{eq:fx}) with $N=1000$,
leading to a continuous but non-differentiable function (in the
limit $N \to +\infty$). Notice that these two functions capture
quite well the complex structure of ups and downs occurring at
multiple scales but fall short of fitting the five sharp drops of
crashes analyzed in the previous Sec.~\ref{S2:Mimick}. Freeing
parameter $\omega$ and adding one parameter $a$ in the fitting
procedure of the quadratic model $\psi_n=a n^2$ improves the
goodness-of-fit by decreasing r.m.s. of the fit residuals to
$\chi=0.274$ with parameters moved to $t_c = {\rm{August~3,
2000}}$, $m = 0.59$, $\omega = 11.2$, $a = 454$, $A = 7.355$, $B =
-0.0090$, and $C = 0.0016$. But again, no sharp drops are
reconstructed due to the incoherent interferences between the
oscillations at different scales.

\begin{figure}
\begin{center}
\epsfig{file=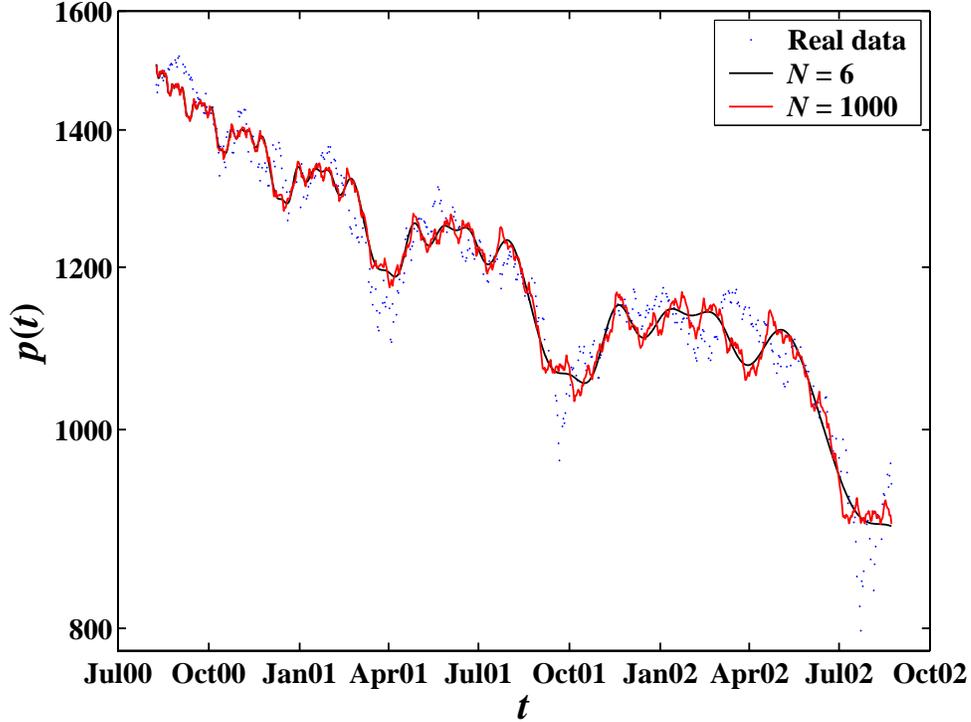,width=13cm}
\end{center}
\caption{Fit of the price trajectory of the US S\&P 500 (dots)
using Eq.~(\ref{eq:fx}) with $N=6$ and $\psi_n=\omega n \ln(\omega
n)$. The values of the parameters are: $t_c= {\rm{July~27,
2000}}$, $m=0.67$, $\omega=11.18$, $A=7.34$, $B=-0.0051$, and $C=
0.0009$. The r.m.s. of the fit residuals is $0.0309$. The thin
wiggly line is constructed by inserting the fitted parameters into
Eq.~(\ref{eq:fx}) with $N=1000$. Freeing the parameter $\omega$ in
the phase $\psi_n=\omega n \ln(\omega n)$ by adopting $\psi_n = b
n \ln(c n)$ improves the goodness-of-fit to the r.m.s. of fit
residuals equal to $\chi=0.280$.} \label{FigWeiWnlnWn}
\end{figure}

As a second illustrative example, let us take the phases
corresponding to the asymptotic dependence of the phases of the
Weierstrass function given by (\ref{eq:Ancos}), that is, $\psi_n =
\omega n \ln(\omega n)$. With $N=6$, we follow the same procedure
as in Sec.~\ref{S2:Mimick}. Fig.~\ref{FigWeiWnlnWn} shows the
fitted function (smooth continuous line) with parameters $t_c=
{\rm{Jul~27, 2000}}$, $m=0.67$, $\omega=11.18$, $A=7.34$,
$B=-0.0051$, and $C= 0.0009$. The r.m.s. of the fit residuals is
$0.0309$. The wiggly continuous curve is constructed by inserting
these fitted parameters into Eq.~(\ref{eq:fx}) with $N=1000$,
leading to a continuous but non-differentiable function (in the
limit $N \to +\infty$). Again, this model does not capture the
five crashes described in Sec.~\ref{S2:Mimick}. Replacing $\omega$
by two free parameters to obtain $\psi_n = b n \ln(c n)$ improves
the goodness-of-fit by decreasing the r.m.s. of the fit residuals
to $\chi=0.280$, but without much differences in the other
parameters and no improvement to describe the 5 crashes.

In sum, Fig.~\ref{FigWeiWn2} and Fig.~\ref{FigWeiWnlnWn} put in
contrast with Fig.~\ref{FigWeiphase0} show that the five crashes
can be described accurately together with the overall anti-bubble
structure by a coherent interference of the log-periodic harmonics
occurring for the model with $\psi_n=a_1n+a_0$ while any other
model of phases that desynchronizes the harmonics destroys the
singularities and leads to a larger r.m.s. of the fit residuals.

\section{Predictions}
\label{S:predict}

Up to now, we have shown that the model (\ref{eq:fx}) with
$\psi_n=0$ provides the best parsimonious description of the
anti-bubble regime of the USA S\&P500 index, improving upon our
previous model \cite{SZ02} by fitting remarkably well the five
crashes indicated by the arrows in Fig.~\ref{FigWeiPrice}.
Notwithstanding the justification of our approach by the general
renormalization group model, we do not have a microscopic model
allowing us to justify rigorously the model (\ref{eq:fx}) with
$\psi_n=0$ based upon the detailed understanding of how each
individual investor behavior is renormalized into a collective
dynamics of this type. As an alternative, we propose that a
crucial test of the model (\ref{eq:fx}) with $\psi_n=0$ lies in
its predictive power. We thus analyze its retro-active predictive
power and then present a prediction for the future, that refines
and complement our prediction issued in \cite{SZ02}.

\subsection{Retroactive predictions} \label{S2:Postdict}

By retroactive predictions, we mimick a real-life situation in
which, at any present time $t_{\rm{last}}$, we fit the past time
series up to $t_{\rm{last}}$, issue a prediction on the evolution
of the price over, say, the next month and then compare it with the
realized price from $t_{\rm{last}}$ to one month after
$t_{\rm{last}}$.

\begin{figure}
\begin{center}
\epsfig{file=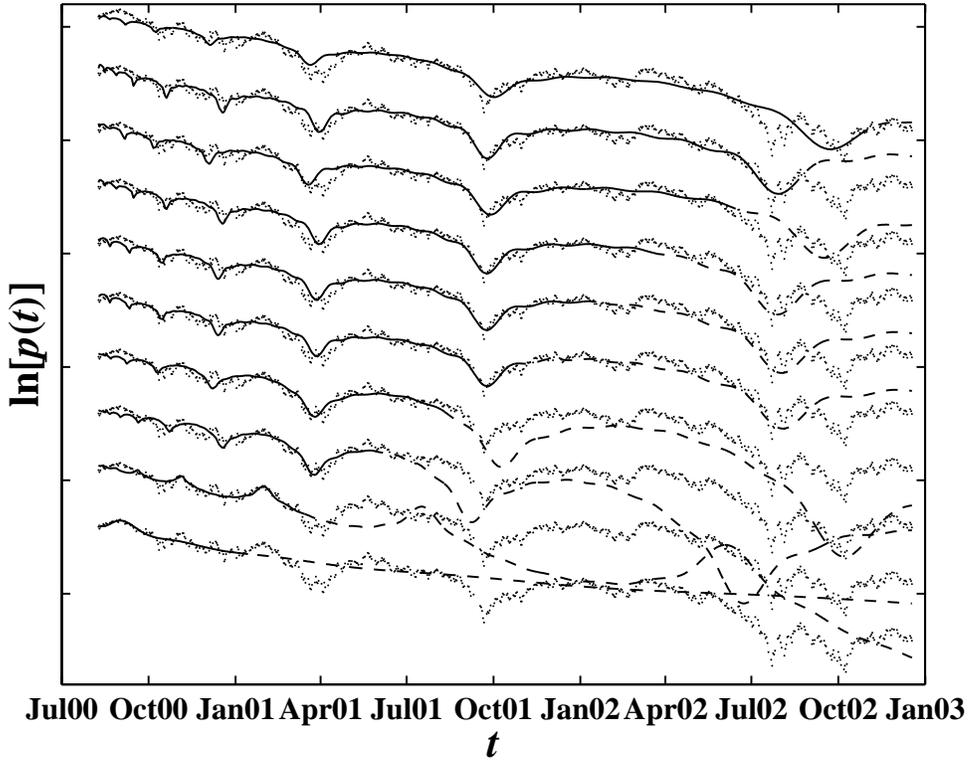,width=13cm}
\end{center}
\caption{Ten predictions of the S\&P 500 index using
Eq.~(\ref{eq:fx}) with $N=6$ and $\psi_n=0$ for ten fictitious
present dates: 02-Jan-2001, 15-Mar-2001, 25-May-2001, 07-Aug-2001,
23-Oct-2001, 04-Jan-2002, 19-Mar-2002, 30-May-2002, 12-Aug-2002,
and 22-Oct-2002. Ten replicas of the price trajectory of the US
S\&P 500 (dots) have been generated and translated vertically to
allow a simple comparison of the predictions. The continuous
curves correspond to the fitted part of the mathematical function
Eq.~(\ref{eq:fx}) with $N=6$ and $\psi_n=0$. The dashed parts give
the extrapolations of these function beyond the fictitious present
dates and thus correspond to the predictions.}
\label{FigWeiPostdict00}
\end{figure}

Figure~\ref{FigWeiPostdict00} compares the realized future prices
to ten predictions of the S\&P 500 index using Eq.~(\ref{eq:fx})
with $N=6$ and $\psi_n=0$ for ten fictitious present dates from
Jan 02, 2001 (bottom) to Oct 22, 2002 (top). Showing ten
translated replicas of the S\&P 500 allows us to represent and
compare the ten predictions in a synoptic manner. In each replica,
the dots are the real price data, the smooth continuous line shows
the fit up to the fictitious present $t_{\rm{last}}$ and the
dashed line is the prediction beyond $t_{\rm{last}}$ obtained by
extrapolating Eq.~(\ref{eq:fx}).

The first two predictions for $t_{\rm{last}}={\rm{Jan~2, 2001}}$
and $t_{\rm{last}}={\rm{Mar~15, 2001}}$ fail to give the correct
trend. The predictions for $t_{\rm{last}}={\rm{May~25, 2001}}$ and
$t_{\rm{last}}={\rm{Aug~7, 2001}}$ capture the trend of the price
trajectory but are too pessimistic. However, these two predictions
suggest that there is a significant endogenous drop in the
beginning of September, 2001, before the 9/11 terrorist attack
happened. In general, the predictions after $t_{\rm{last}} =
{\rm{Oct~23, 2001}}$ are in excellent agreement with the real data
and are very robust. In addition, the last three predictions are
successful in suggesting the future happening of an instability in
the form of two close sharp drops between July and November of
2002.

\begin{figure}
\begin{center}
\epsfig{file=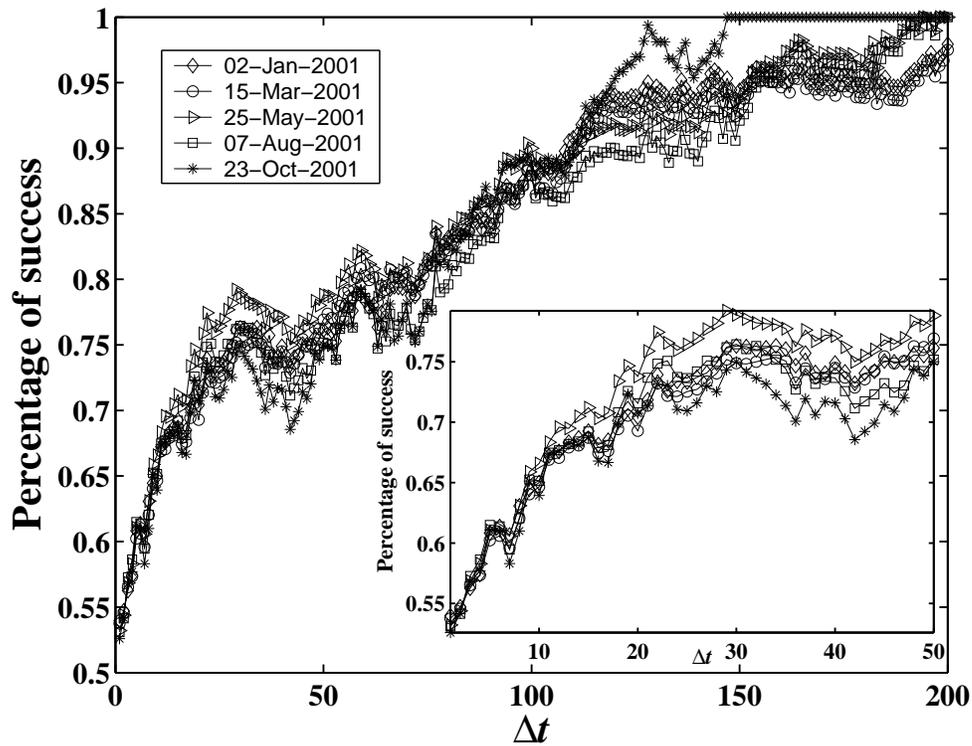,width=13cm}
\end{center}
\caption{Percentage of success for the prediction of the signs of
the returns over a time horizon of $\Delta t$ as a function of
$\Delta t$ for five different starting dates of the strategies
given in the upper-left inset. The lower-right inset gives a
magnification for $\Delta t \le 50$. See text for explanations.}
\label{FigWeiPostdict01}
\end{figure}

Let us now quantify the predictive skill of Eq.~(\ref{eq:fx}) with
$N=6$ and $\psi_n=0$ by calculating the percentage of success for
the prediction of the sign of the return $r(t; \Delta t) =
\ln[p(t+\Delta t)] - \ln[p(t)]$ from the fictitious present
$t_{\rm{last}}$ to the time $t_{\rm{last}} + \Delta t$, that is,
over a time horizon of $\Delta t$ days. For each successive
fictitious present $t_{\rm{last}}$, a new prediction is issued to
predict the sign of $r(_{\rm{last}}; \Delta t)$. We investigate
prediction horizons $\Delta t$ spanning from $1$ day to $200$ days.
For each $\Delta t$, we calculate the success rate from a starting
time $t_{\rm{enter}}$ to November, 21, 2002 and consider five
different starting time $t_{\rm{enter}}$ to test for the
robustness of the results. The results are presented in
Fig.~\ref{FigWeiPostdict01}. The prediction success percentage
increases with $\Delta t$ from a minimum of $53\%$ for $\Delta t =
1$ day to above $90-95\%$ for $\Delta t > 120$ days. This
extraordinary good success for the larger time horizons only
reflects the overall downward trend of the S\&P500 market over the
studied time period and is thus not a qualifier of the model. More
interesting is the observation that the success rate increases
above $70\%$ for $\Delta t > 20$ days. In addition, the results
are robust with respect to the changes of the starting date
$t_{\rm{enter}}$. We consider the plateaus at about $75\%$ success
rate in the range $\Delta = 20 \sim 70$ days to reflect a genuine
predictive power of the model.

If indeed the model has predictive power, we should be able to use
it in order to generate a profitable investment strategy. This is
done as follows. For simplicity, we assume that a trader can hold
no more than one share (position $+1$) or short no more than one
share (position $-1$). In addition, a trader is always invested in the market,
so that her position is $\pm 1$ at any time. For a given $\Delta
t$, at a given time $t_{\rm{last}}$, if the model predicts that,
at time $t_{\rm{last}} + \Delta t$, the price will go up
($r(t_{\rm{last}};\Delta t)>0$), the trader buys 2 shares at the
price $p(t_{\rm{last}})$ if she was holding $-1$ share or holds
her existing share if she has already one. Analogously, if the
prediction shows that the price will go down
($r(t_{\rm{last}};\Delta t)<0$), she sells 2 shares at the present
time $t_{\rm{last}}$ at the price $p(t_{\rm{last}})$ if she was
holding $1$ share or holds her existing $-1$ short position if she
is shorting already. Each prediction and reassessment of her
position is performed with a time step of $\Delta t$ days. We do
not include transaction costs or losses from the motion of the
price during a transaction. Their impact can easily be
incorporated and they are found not to erase the significant
arbitrage opportunity documented here.

\begin{figure}
\begin{center}
\epsfig{file=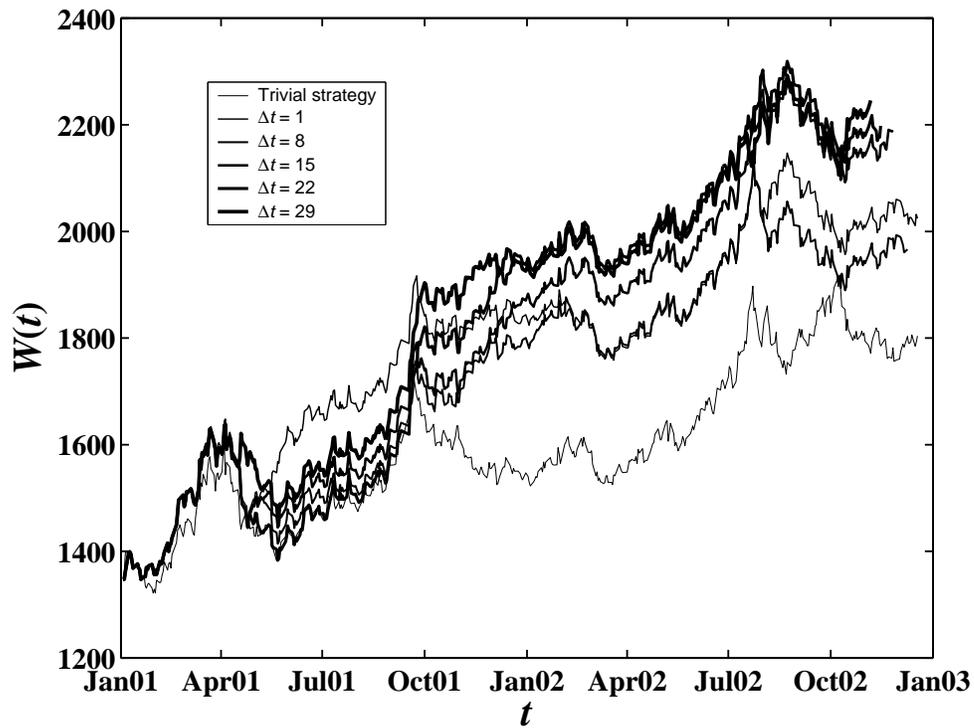,width=13cm}
\end{center}
\caption{The running wealth $W(t)$ of a trader investing with a
time step $\Delta t$ based on the prediction of the sign of the
return over the time interval $\Delta t$. The origin of time is
chosen as $t_{\rm{enter}}={\rm{Jan~2, 2001}}$. Six time horizons
are compared with the ``short-and-hold'' strategy (called trivial
in the inset). See text for explanations.} \label{FigWei_Wt}
\end{figure}

\begin{figure}
\begin{center}
\epsfig{file=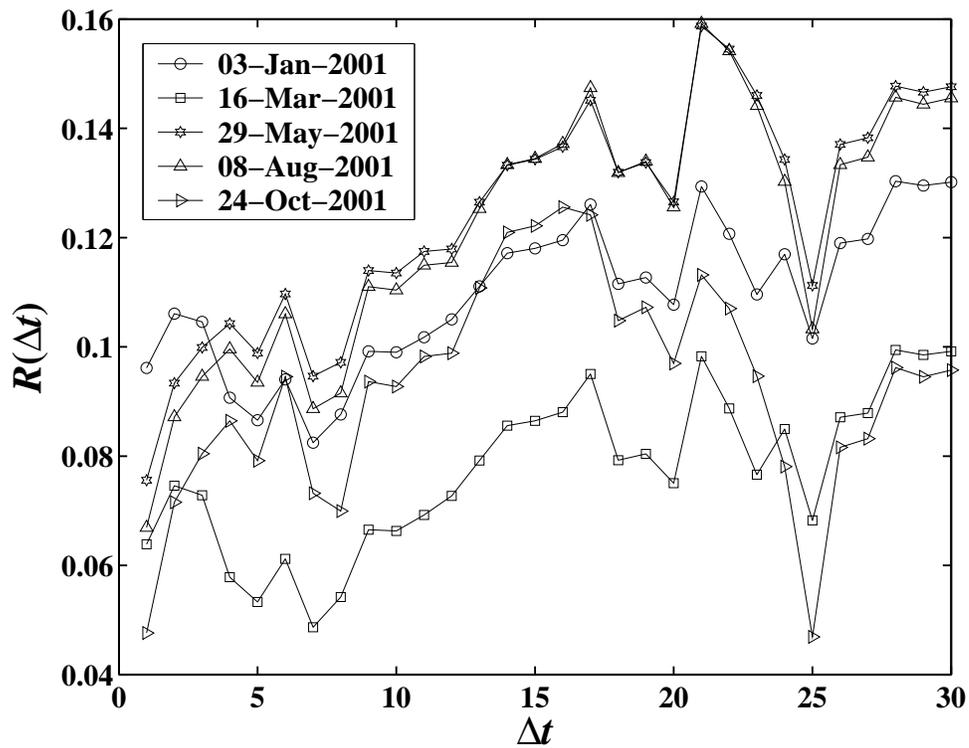,width=13cm}
\end{center}
\caption{Daily Sharpe ratio $R(\Delta t)$ as a function of the
time horizon $\Delta t$ for five different origin of time
$t_{\rm{enter}}$.} \label{FigWei_R}
\end{figure}

The initial wealth of each trader is $+1$ share of stock at the
initial time taken to be $t_{\rm{enter}}={\rm{Jan~2, 2001}}$.
Figure \ref{FigWei_Wt} shows the time dependence of the trader's
wealth $W(t)$ for different values of $\Delta t$. We compare this
wealth with the ``short-and-hold'' strategy, the symmetric of the
usual ``buy-and-hold'' strategy adapted to the overall bearish
nature of the S\&P500 price trajectory over the studied period. It
is clear that all the strategies using our model outperform
significantly the naive ``short-and-hold'' strategy. The
corresponding Sharpe ratios, $R(\Delta t)$, defined as the ratio
of the average daily return $\ln[W(t+1)] - \ln[W(t)]$ divided by
the standard deviation of the daily return, are shown in Figure
\ref{FigWei_R} as a function of $\Delta t$ for five different
starting times $t_{\rm{enter}}$ of the strategies. Our trading
strategies have a daily Sharpe ratio of the order of $0.1-1.4$. In
comparison, the Sharpe ratio of the ``short-and-hold'' strategy is
$0.06$. Thus, the impact of our strategy is roughly to double the Sharpe
ratio. Translated into the standard yearly sharp ratio, a daily value
$0.1-0.14$
gives a quite honorable value in the range $1.5-2.2$ compared with
$0.9$ for the ``short-and-hold.''

\subsection{Forward predictions} \label{S2:Forward}

\begin{figure}
\begin{center}
\epsfig{file=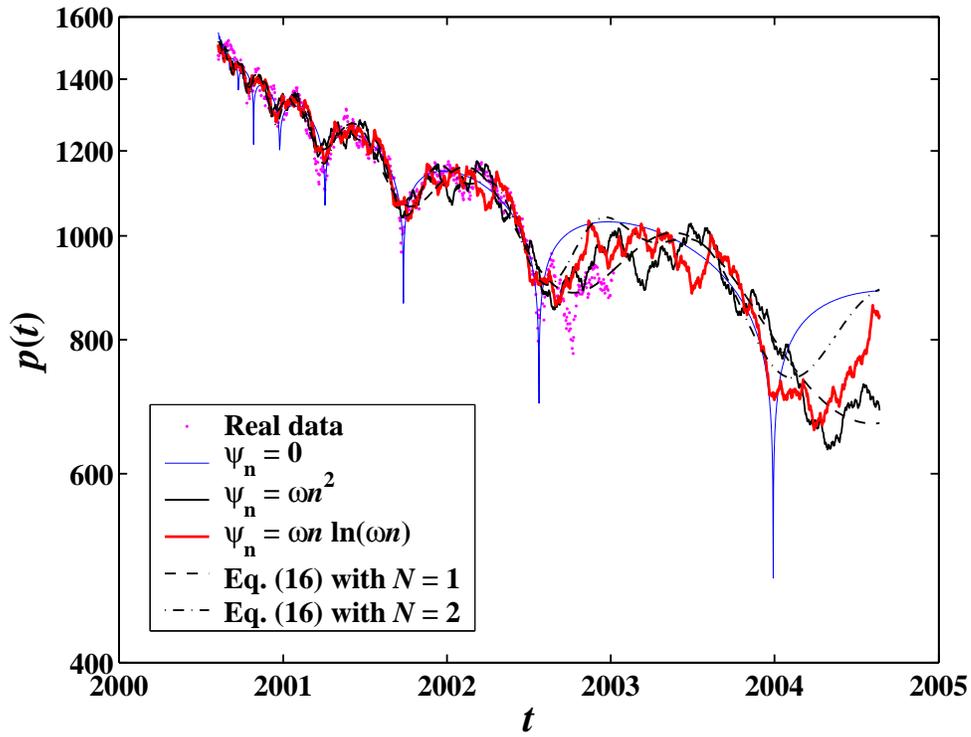,width=13cm}
\end{center}
\caption{Forward predictions based on the analysis of the US S\&P
500 index from August 21, 2000 to August 24, 2002 using
Eq.~(\ref{eq:fx}) with $N=1000$ obtained with the three phase
models $\psi_n$ indicated in the legend. The otherwise lines are
predictions using (\ref{Eq:lnpt}) with $N=1$ (dashed line) and
$N=2$ (dash-dotted line).} \label{FigWeiPredict01}
\end{figure}

The predictions shown in Fig.~\ref{FigWeiPredict01} use the US
S\&P 500 index from August 21, 2000 to August 24, 2002 which is
fitted by Eq.~(\ref{eq:fx}) with $N=6$ and three different phases
$\psi_n=0$, $\psi_n=\omega n^2$ and $\psi_n = \omega n \ln(\omega
n)$ respectively. Once the parameters of these three models have
been determined, we construct the corresponding Weierstrass-type
functions with $N=1000$ terms. The predictions are obtained as
straightforward extrapolations of the three Weierstrass-type
functions in the future. The predictions published in \cite{SZ02}
using the log-periodic formula (\ref{Eq:fit1}) with a single
angular log-frequency $\omega$ (dashed line) and including its
harmonics $2 \omega$ (dash-dotted line) are also shown for
comparison.

\begin{figure}
\begin{center}
\epsfig{file=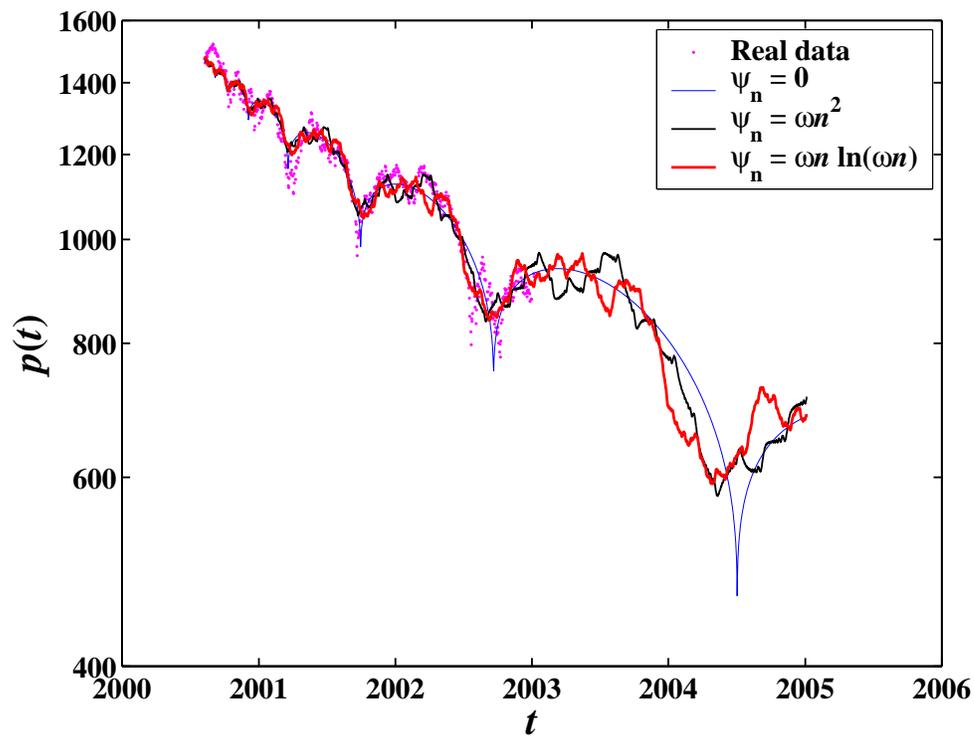,width=13cm}
\end{center}
\caption{Same as Figure \ref{FigWeiPredict01} based on the
analysis of the US S\&P 500 index from August 21, 2000 to January
8, 2003.} \label{FigWeiPredict02}
\end{figure}

Fig.~\ref{FigWeiPredict02} is the same as
Fig.~\ref{FigWeiPredict01} but is updated by including the US S\&P
500 index data from August 21, 2000 to January 8, 2003. One can
observe that the two predictions are very similar. The parameters
of the corresponding fits are listed in Table~\ref{Tb1}.

According to these predictions, we can expect an overall
continuation of the bearish phase, punctuated by some local
rallies. An overall increasing market until the first quarter of
2003 is predicted, which should be followed by several months of
more or less stable prices. Then, we predict a rather rapid
descent (with maybe one or two severe ups and downs in the middle)
which bottoms during the first semester of 2004. Quantitatively,
the predictions suggests that the S\&P 500 will culminate at about
1000 in the first quarter of 2003 and then dive down to 650 in
early/mid 2004. These results are consistent with those in
Ref.~\cite{SZ02}.

In spite of the robustness of the predictions, there is nevertheless
a slight visible differences between Fig.~\ref{FigWeiPredict01} and
Fig.~\ref{FigWeiPredict02}. The latest predictions
shown in Fig.~\ref{FigWeiPredict02} are shifted
compared with those shown in Fig.~\ref{FigWeiPredict01} by a few months.
Technically, we see that this comes from the existence of two dips in the second
half of 2002, the first dip being fitted by the model in
Fig.~\ref{FigWeiPredict01} while the second dip seems to have a stronger
influence as shown in Fig.~\ref{FigWeiPredict02}.

\section{Discussion and Concluding remarks}
\label{S:concl}

We have proposed a straightforward extension of our previously
proposed log-periodic power law model of the ``anti-bubble''
regime of the USA market since the summer of 2000 \cite{SZ02} and
which is continuing to the present day. The generalization
proposed here is using the renormalization group framework to
model critical points. Using a previous work \cite{GS02} on the
classification of the class of Weierstrass-like function, we have
shown that the five crashes that occurred since August 2000 can be
accurately modelled by this approach, in a fully consistent way
with no additional parameters. Our theory suggests an overall
consistent organization of the investors forming a collective
network which interact to form the pessimistic bearish
``anti-bubble'' regime with intermittent acceleration of the
positive feedbacks of pessimistic sentiment leading to these
crashes. We have complemented the descriptive analysis by
developing and testing retrospective predictions, that confirm the
existence of significant arbitrage opportunities for a trader
using our model. We then conclude our analysis by offering a
prediction for the unknown future. Obviously, only forward
predictions constitute the ultimate proof of our claims. This
exercise is an additional step in our constitution of a database
of forward predictions whose statistical analysis will eventually
allow us to confirm or disprove the validity of our models.

An additional feature may be pointed out: our analysis suggests
that the five crashes are essentially of an endogenous origin,
including the one in the first half of September 2001, often
thought to have been provoked by the 911 terrorist attack. Our
analysis suggests that the market was already on its way to a
panic: the 911 event seems to have amplified the drop
quantitatively without changing its qualitative nature. A similar
conclusion was obtained for the crash of August-September 1998
often attributed to external political and economic events in
Russia \cite{Nikkei99}. In this spirit, Ref.~\cite{EndoExo} has
stressed that two-third of the major crashes in a large number of
markets are of an endogenous origin and only the most dramatic
piece of news (such as the announcement of World War I) are able
to move the markets with an amplitude similar to the endogenous
crashes.

\bigskip
{\bf Acknowledgments:} We are grateful to V.F. Pisarenko for
helpful discussions concerning the Jack-knife method and S.
Gluzman for useful discussions on Weierstrass-type functions. This
work was supported by the James S. Mc Donnell Foundation 21st
century scientist award/studying complex system.


\newpage
\begin{table}
\begin{center}
\caption{\label{Tb1} Values of the parameters of the fits of the
S\&P500 index from Aug-09-2000 to $t_{\rm{last}}$ given in the
first column with formula (\ref{eq:fx}) for different models of
the phases $\psi_n$. $\chi$ is the r.m.s. of the fit residuals.}
\medskip
\begin{tabular}{ccccccccccccc}
 \hline\hline
$t_{\rm{last}}$&$\psi_n$&$t_c$&$m$&$\omega$&$A$&$B$&$C$&$\chi$\\\hline
Aug-24-2002&0&19-Jul-2000&0.52&11.41&7.40&-0.0146&-0.0022&0.027\\
Aug-24-2002&$\omega n^2$&02-Aug-2000&0.69&10.59&7.33&-0.0044&0.0008&0.031\\
Aug-24-2002&$\omega n\ln(\omega n)$&27-Jul-2000&0.67&11.18&7.34&-0.0052&0.0010&0.030\\
Jan-08-2003&0&30-Jul-2000&0.82&10.37&7.31&-0.0019&-0.0004&0.036\\
Jan-08-2003&$\omega n^2$&31-Jul-2000&0.84&10.55&7.31&-0.0016&0.0003&0.035\\
Jan-08-2003&$\omega n\ln(\omega n)$&22-Jul-2000&0.83&11.07&7.31&-0.0017&0.0003&0.033\\
 \hline\hline
\end{tabular}
\end{center}
\end{table}

\clearpage

\end{document}